\documentclass[aps,prd,amssymb,superscriptaddress,nofootinbib,twocolumn,preprintnumbers]{revtex4-2}
\usepackage[utf8]{inputenc}
\usepackage[english]{babel}
\usepackage{amsmath}
\usepackage{amsfonts}
\usepackage{amssymb}
\usepackage{amsthm}
\usepackage{color}
\usepackage{graphicx}
\usepackage[force]{feynmp-auto}
\usepackage{feynmp}
\usepackage[hidelinks]{hyperref}

\newcommand\bw{\begin{widetext}}
\newcommand\ew{\end{widetext}}

 \def\be{\begin{equation}}
\def\ee{\end{equation}}
 \def\ba{\begin{align}}
\def\ea{\end{align}}
\def\bea{\begin{eqnarray}}
\def\eea{\end{eqnarray}}

\def\m{\mu}

\begin{document}
\preprint{FTUAM-20-26}
\preprint{IFT-UAM/CSIC-20-162}
\preprint{INR-TH-2020-043}
\title{{\bf Massless Positivity in Graviton Exchange}}
\author{Mario Herrero-Valea}
\email[]{mherrero@sissa.it}
\address{SISSA, Via Bonomea 265, 34136 Trieste, Italy and INFN Sezione di Trieste}
\address{IFPU - Institute for Fundamental Physics of the Universe \\Via Beirut 2, 34014 Trieste, Italy}

\author{Raquel Santos-Garcia}
\email[]{raquel.santosg@uam.es}
\address{Departamento de F\'isica Te\'orica and Instituto de F\'isica Te\'orica, IFT-UAM/CSIC,
Universidad Aut\'onoma de Madrid, Ciudad Universitaria de Cantoblanco, 28049 Madrid, Spain}

\author{Anna Tokareva}
\email[]{tokareva@ms2.inr.ac.ru}
\address{Department of Physics, P.O.Box 35 (YFL), FIN-40014 Jyv\"askyl\"a\\ University of Jyv\"askyl\"a, Finland}
\address{Institute for Nuclear Research of Russian Academy of Sciences\\
117312 Moscow, Russia}
\address{Helsinki Institute of Physics (HIP),
P.O. Box 64, 00014 Helsinki\\ University of Helsinki,  Finland}

\begin{abstract}
We formulate Positivity Bounds for scattering amplitudes including exchange of massless particles. We generalize the standard construction through dispersion relations to include the presence of a branch cut along the real axis in the complex plane for the Maldestam variable $s$. In general, validity of these bounds require the cancellation of divergences in the forward limit of the amplitude, proportional to $t^{-1}$ and $\log(t)$. We show that this is possible in the case of gravitons if one assumes a Regge behavior of the amplitude at high energies below the Planck scale, as previously suggested in the literature, and that the concrete UV behaviour of the amplitude is uniquely determined by the structure of IR divergences. We thus extend previous results by including a sub-leading logarithmic term, which we show to be universal. The bounds that we present here have the potential of constraining very general models of modified gravity and EFTs of matter coupled to gravitation.
\end{abstract}

\maketitle
\section{Introduction}
Positivity bounds \cite{Nicolis:2009qm,Adams:2006sv,Bellazzini:2017fep,deRham:2018qqo,deRham:2017zjm} have become standard tools in assessing the validity of low energy Effective Field Theories (EFT). By invoking the plausible existence of an ultra-violet (UV) completion satisfying reasonable properties such as Lorentz Invariance, unitarity, and locality, positivity bounds exclude large regions of the parameter space of a given EFT by demanding the positivity of a certain combination of couplings.

In particular, these bounds are obtained by combining the knowledge of the analytic structure of $2$-to-$2$ scattering amplitudes with the optical theorem
\begin{align}\label{eq:optical}
    {\rm Im}{\cal A}(s,0)=s\sqrt{1-\frac{4m^2}{s}}\ \sigma(s),
\end{align}
which ensures positivity of the imaginary part of the scattering amplitude ${\cal A}(s,t)$ in the forward limit $t\rightarrow 0$.

Applications of positivity bounds include the proof of the a-theorem \cite{Komargodski:2011vj,Luty:2012ww}, the study of chiral perturbation theory \cite{Manohar:2008tc}, effective Higgs models \cite{Low:2009di}, Quantum Gravity \cite{Bellazzini:2015cra,AccettulliHuber:2020oou}, massive gravity and Galileons \cite{Cheung:2016yqr,Bonifacio:2016wcb,Keltner:2015xda,deRham:2017imi,Falkowski:2020mjq}, higher spins \cite{Bellazzini:2019bzh}, Cosmology \cite{Baumann:2015nta,Croon:2015fza,Melville:2019wyy,Herrero-Valea:2019hde}, String Theory \cite{Chen:2019qvr,Green:2019tpt}, and many more. Recently, a generalization of positivity bounds, named \emph{arcs}, has been proposed \cite{Bellazzini:2020cot}.

However, all these examples omit an important case of physical relevance, the exchange of massless particles. In that case, the scattering amplitude contains pathologies that impede to take the forward limit -- a pole $t^{-1}$ and a logarithmic divergence $\log(t)$, due to exchange and production of massless particles. This is particularly relevant in the presence of Gravity, since gravitons couple to all forms of matter. Although for energies below the Planck scale gravity could be ignored, its character as a long range force produces contributions to the scattering amplitude down to the deep infra-red (IR). Formally, the pathologies which come with the exchange of gravitons are never absent and cast a shadow on the validity of positivity bounds. Even if one trusts the decoupling limit and the validity of gravity-less positivity bounds, it would be desirable to find a way to extend them to include graviton exchange. There have been previous attempts to solve this issue by compactifying space-time down to three dimensions, where gravitons decompose in massive fields \cite{Bellazzini:2019xts,Alberte:2020jsk}, but a general formalism applicable in more varied situations, without requiring compactification, is still lacking.

Recently, it has been suggested that forward divergences in graviton exchange can be cancelled by assuming  a Regge form for the high-energy limit of the scattering amplitude \cite{Tokuda:2020mlf}, which is expected to hold from String Theory \cite{Camanho:2014apa,DAppollonio:2015fly}. However, in \cite{Tokuda:2020mlf} only the term $t^{-1}$ is cancelled and nothing is said about the logarithm. This is important, though, because due to crossing symmetry, equivalent $\log(s)$ and $\log(u)$ terms are expected to co-exist in the scattering amplitude. These terms split the complex plane in $s$ in two, with a branch cut along the real line for $t\rightarrow 0$. This obstructs the formulation of usual positivity bounds, which require to deform an integration contour crossing the real axis.

In this \emph{paper} we construct \emph{new} positivity bounds for theories with exchange of massless particles, provided that we cancel the divergences in the forward limit. We show that this is indeed possible when the massless states correspond to gravitons. Generalizing the results of \cite{Tokuda:2020mlf}, we prove that both the pole $t^{-1}$ and the $\log(t)$ can be eliminated, with the remaining pieces in the amplitude satisfying a positivity bound reminiscent of the standard case. Finally, we discuss the robustness of our result by showing agreement with previous works in the literature, formally deriving the bounds recently proposed by \cite{Bellazzini:2019xts,Alberte:2020jsk}.
\\
\section{Dispersion Relations}
From now on we will consider $ab\rightarrow ab$ scattering amplitudes which include a massless particle coupled to the bosonic external states $a$ and $b$. The presence of this massless state will produce poles in $s$, $t$ and $u$ from tree-level exchange, as well as logarithmic cuts $\log(s)$, $\log(t)$ and $\log(u)$ indicating particle production, found at loop level in perturbation theory. Here $s$, $t$ and $u$ are the Maldestam variables, with $s$ the energy in the center of mass frame squared. $u$ can always be eliminated by using $s+t+u=4m^2$, where we have assumed that both states $a$ and $b$ have the same mass $m$. From Cauchy's integral theorem, one can write a family of dispersion relations for the amplitude
\begin{align}\label{eq:disp}
    {\cal A}(s,t)=\frac{(s-\mu)^n}{2\pi i}\oint_{\gamma_s}dz\ \frac{{\cal A}(z,t)}{(z-s)(z-\mu)^n},
\end{align}
with $n\geq 1$. The integration contour $\gamma_s$ must be taken as a small circle surrounding only the point $z=s$, while the point $z=\mu$ is arbitrary provided that it lays outside the contour. We take $\mu$ real hereinafter.

A key point in deriving positivity bounds lays on the behavior of the scattering amplitude at high energies. For massive particles, it can be proven that it satisfies the Froissart-Martin bound \cite{Martin:1965jj}, which implies
\begin{align}\label{eq:frm}
    \lim_{|s|\rightarrow \infty}\left|\frac{{\cal A}(s,t)}{s^2} \right|=0, \quad t<4m^2.
\end{align}

Alas, the formal proof of this bound cannot be applied to exchange of massless particles. However, we will assume that this is still true for the cases considered here. We will justify this assumption later.

Now we take the forward limit of \eqref{eq:disp}. In the case of massless particles in the intermediate channel, this is divergent and cannot be taken exactly. We thus instead, in more generality\footnote{This form encodes all the cases of relevance to our knowledge. For exchange of scalars and vectors, both $f(s)$ and $g(s)$ are constant, while for gravitons, they behave as $\sim s^2$.}, expand the amplitude around the limit $t\rightarrow 0^-$
\begin{align}\label{eq:exp_amp}
    \nonumber &{\cal A}(s,0^-)\equiv \left. {\cal A}(s,t)\right|_{t\rightarrow 0^-}\\
    &=\frac{f(s)}{t}+g(s)\log(t)+{\cal A}_\circ (s)+{\cal O}(t),
\end{align}
where the limit is taken from the negative side of the real line.

Here $f(s)$ and $g(s)$ are holomorphic functions. When the scattering amplitude is computed in perturbation theory, $f(s)$ contains the residue on the pole of the massless propagator, while $g(s)$ is proportional to the $\beta$-function of the $ab\rightarrow ab$ coupling. The analytic structure of ${\cal A}(s,0^-)$ is therefore controlled by ${\cal A}_\circ(s)$. This is analytic in the whole complex plane except for a branch cut running over the whole real line, due to production of massless particles and crossing symmetry \cite{Martin:1965jj}.

\begin{figure}
  \includegraphics[scale=.25]{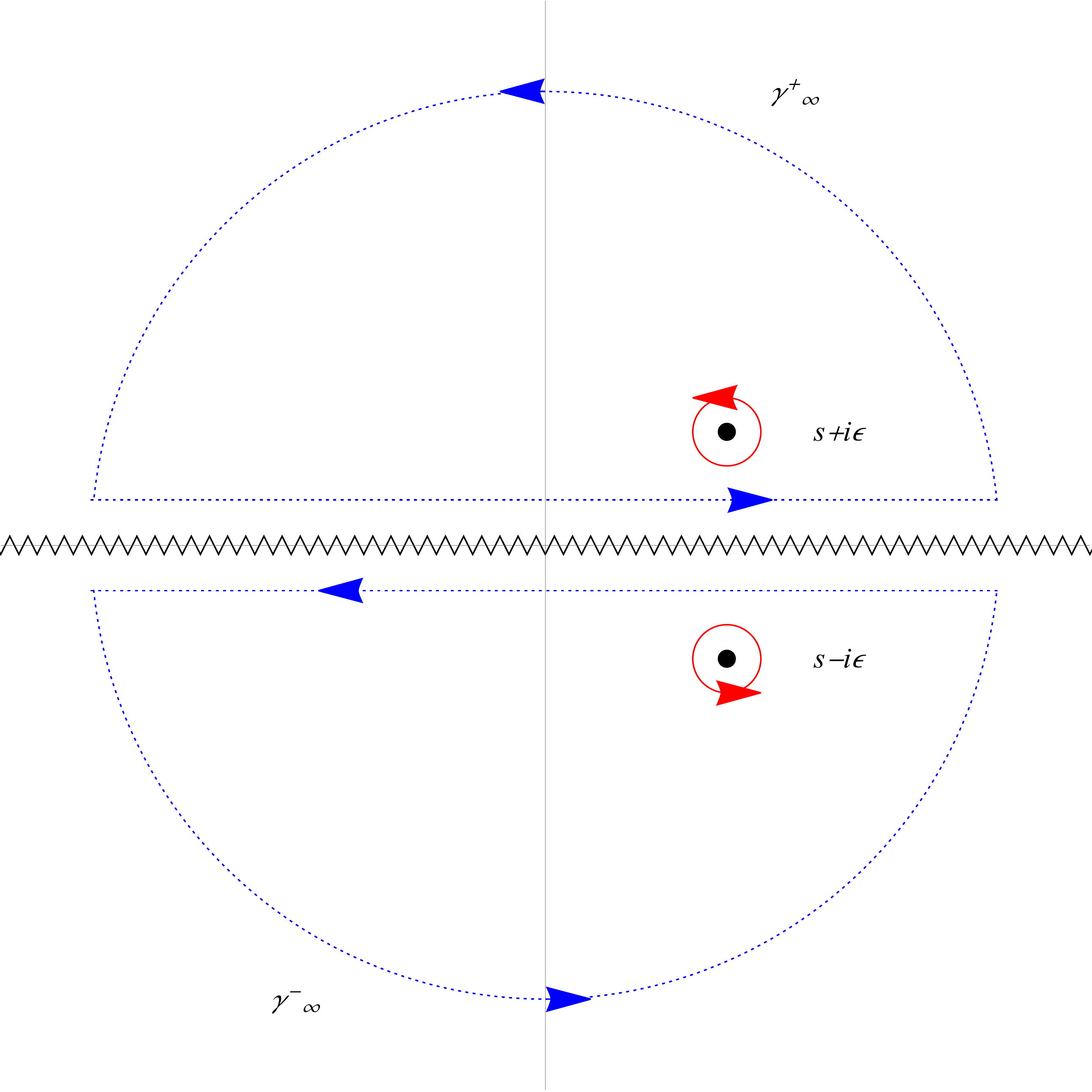}
\caption{Integration contours in the complex plane for $s$. The zigzag line represents the branch cut. For points $s\pm i\epsilon$, the integration contour $\gamma_s$ in the corresponding half of the complex plane is shown in red. The equivalent contours used in \eqref{eq:newint} are dotted in blue. The radius of the large semi-circumferences $\gamma_\infty^\pm$ is $|s|\rightarrow \infty.$}
\label{fig:contours}
\end{figure}

The branch cut obstructs the standard derivation of positivity bounds, which uses a contour integral crossing the real line \cite{Adams:2006sv}. Here instead we note that for any real value of $s$, we can perform two different analytic continuations of the amplitude, by adding a small imaginary part $s\pm i\epsilon$ which moves the point to the upper (down) part of the complex plane. Afterwards we can deform the integration contour to run above (below) the real axis plus a semi-circumference at infinity, as shown in figure \ref{fig:contours}. This allows to define two different realisations of \eqref{eq:disp} 
\begin{align}\label{eq:newint}
&{\cal A} (s+i \epsilon,0^-) =\frac{(s -\mu)^n}{2\pi i}\int_{-\infty}^\infty dz\ \frac{{\cal A}(z+i \epsilon,0^-)}{(z-s)(z-\mu)^n},\\
&{\cal A}(s- i\epsilon,0^-) =\frac{(s-\mu)^n}{2\pi i}\int_{\infty}^{-\infty} dz\ \frac{{\cal A}(z-i \epsilon,0^-)}{(z-s)(z-\mu)^n},
\end{align}
where the circles at infinity vanish due to \eqref{eq:frm} and $\epsilon$ must be understood as infinitesimal. Here we keep it finite only on non-holomorphic tems. Subtracting both representations we get
\begin{align}
\nonumber &{\cal A}(s+ i \epsilon,0^-)-{\cal A} (s-i\epsilon,0^-)\\
&=\frac{(s-\mu)^n}{2\pi i}\int_{-\infty}^{\infty}dz\ \frac{{\cal A}(z+i \epsilon,0^-)+{\cal A}(z-i \epsilon,0^-)}{(z-s)(z-\mu)^n}.
\end{align}

In the physical region $s\in \mathbb{R}$ and we have ${\cal A}^*(s-i\epsilon,t)={\cal A}(s+i\epsilon,t)$. Then
\begin{align}
{\rm Im}{\cal A}(s\pm i\epsilon,0^-)=\mp \frac{(s-\mu)^n}{2\pi}\int_{-\infty}^{\infty}dz\ \frac{{\rm Re}{\cal A}(z\pm i \epsilon,0^-)}{(z-s)(z-\mu)^n}.
\end{align}

We now take ${\cal A} (s+ i \epsilon,0^-)$ and use this result to rewrite it as
\begin{align}\label{eq:dispB}
\nonumber &{\cal A}(s+i\epsilon,0^-)-i\  {\rm Im}{\cal A}(s+i \epsilon,0^-)\\
&={\rm Re}{\cal A}(s+i\epsilon,0^-)=\frac{(s-\mu)^n}{2\pi }\int_{-\infty}^\infty dz\ \frac{{\rm Im}{\cal A}(z+i \epsilon,0^-)}{(z-s)(z-\mu)^n}.
\end{align}

This expression is reminiscent of the standard derivation of positivity bounds. However, in our case we have cancelled out the imaginary part of the amplitude, getting rid of the discontinuity explicitly.

The integral in \eqref{eq:dispB} runs over non-physical values of $z$. This can be solved by splitting it in three integrals over $\{-\infty,0\}$, $\{0,4m^2 \}$ and $\{4m^2,\infty\}$. Performing a change of variables $z\rightarrow -z+4m^2$ in the first one and using crossing symmetry, \eqref{eq:dispB} can be rewritten as
\begin{widetext}
\begin{align}\label{eq:B2}
    &{\cal B}(s,0^-)=\frac{(s-\mu)^n}{2\pi}\int_{4m^2}^\infty dz\ \left(
  \frac{{\rm Im}{\cal A}(z+ i\epsilon,0^-)}{(z-s)(z-\mu)^n}+\frac{(-1)^n{\rm Im}{\cal A}^{\times}(z+ i\epsilon,0^-)}{(z-4m^2+s)(z-4m^2+\mu)^n}\right),
\end{align}
\end{widetext}
where we have defined
\begin{align}\label{eq:defB}
    \nonumber &{\cal B}(s,0^-)={\rm Re}{\cal A}(s+i\epsilon,0^-)\\
    &-\frac{(s-\mu)^n}{2\pi }\int_{0}^{4m^2} dz\ \frac{{\rm Im}{\cal A}(z+i \epsilon,0^-)}{(z-s)(z-\mu)^n}.
\end{align}
Here ${\cal A}^{\times}(s,0^-)={\cal A}(-s+4m^2,0^-)+{\cal O}(t)$ is the crossed amplitude in the $u$-channel.

Now, by using the optical theorem \eqref{eq:optical} in the rhs of \eqref{eq:B2}, we would be tempted to follow the standard derivation of positivity bounds, and conclude that
\begin{align}
 \nonumber   &\frac{1}{n!}\left. \frac{d^n}{ds^n}{\cal B}(s,0^-)\right|_{s=0}\\
 &=\int_{4m^2}^\infty \frac{dz}{2\pi}\ z\sqrt{1-\frac{4m^2}{z}}\left(
  \frac{\sigma (z)}{z^{n+1}}+\frac{(-1)^n \sigma^\times (z)}{(z-4m^2)^{n+1}}\right)>0,
\end{align}
for even $n$, after taking derivatives in both sides of \eqref{eq:B2}.

Nonetheless, this is not possible in the case at hand. Barring aside the issue of the forward limit divergences -- which we will discuss later --, we must note that the lhs of \eqref{eq:B2} can be IR divergent in perturbation theory. For finite masses, the integral piece in \eqref{eq:defB} takes care of these IR divergences, replacing them by $m^{-2}$ and $\log(m^2)$. However, this will not work in the massless case. To circumvent this issue, we take \eqref{eq:B2} and define instead the following function
\begin{align}\label{eq:sigma}
&\Sigma^{(j)}=\frac{1}{2\pi i}\oint_{\gamma_\delta}ds\ \frac{s^3 {\cal B}(s,0^-)}{(s^2+\delta^2)^{2j+1}}
\end{align}
where now $n=2j$ and $\delta$ has dimensions of energy squared. Using now \eqref{eq:B2} we find
\begin{align}\label{eq:Fdef}
    &\Sigma^{(j)}=\int_{4m^2}^\infty dz\ F^{(j)}(z),
\end{align}
were we have performed the integral in $s$ explicitly, getting
\begin{align}
\label{rhs}
\nonumber   & F^{(j)}(z)=\frac{z^3 {\rm Im}{\cal A}(z+i\epsilon,0^-)}{2\pi (z^2+\delta^2)^{2j+1}}\\
    &+\frac{(z-4m^2)^3{\rm Im}{\cal A}^\times(z+i\epsilon,0^-)}{2\pi((z-4m^2)^2+\delta^2)^{2j+1}}.
\end{align}

Notice that all dependence on $
\mu$ has cancelled after integration, with $
\delta$ taking its place in the denominators. The contour $\gamma_\delta$ is the sum of two small circles enclosing the points $s=\pm i\delta$, with $\delta >0$. It acts as sort of an IR regulator, but it is not constrained to be small. Note also that the rhs of \eqref{eq:Fdef} is positive definite for all $j$, since $F^{j}(z)>0$ within the integration regime from application of \eqref{eq:optical}. In that case, we can conclude that
\begin{align}\label{eq:positivity}
\Sigma^{(j)}=\frac{1}{2\pi i}\oint_{\gamma_\delta}ds\ \frac{s^3 {\cal B}(s,0^-)}{(s^2+\delta^2)^{2j+1}}>0,
\end{align}
regardless of the shape of the scattering amplitude, which might be even unknown above a certain energy scale $\Lambda$.

Indeed, let us assume that ${\cal A}(s,t)$ is known only within an EFT with validity up to\footnote{Note that $\Lambda$ might not be strictly the cut-off of the theory, but the energy at which the EFT is not a good approximation to the UV complete theory anymore. This might happen a few orders of magnitude below the cut-off.} $E\sim \Lambda\gg m,\delta$. In that case, we can split the integral on the rhs and rewrite \eqref{eq:sigma} as
\begin{align}\label{eq:sigmabar}
    \hat{\Sigma}^{(j)} =\int_{\Lambda^2}^\infty dz\ F^{(j)}(z) ,
\end{align}
where
\begin{align}
   \hat{\Sigma}^{(j)}=\Sigma^{(j)}-\int_{4m^2}^{\Lambda^2} dz\ F^{(j)}(z).
\end{align}

Again, the rhs of \eqref{eq:sigmabar} is positive and we conclude
\begin{align}\label{eq:beyond}
    \hat{\Sigma}^{(j)}>0.
\end{align}
Expressions \eqref{eq:positivity} and \eqref{eq:beyond} are the massless version of positivity and \emph{beyond} positivity bounds \cite{Bellazzini:2017fep}. They state that the contour integral in \eqref{eq:sigma} -- or the quantity $\bar{\Sigma}^{(j)}$ in \eqref{eq:sigmabar} --, which can be computed in an EFT provided that it is valid below $\Lambda$, has to be positive. They differ from standard positivity bounds in two manners, which encode the particularities of the massless exchange. First, we find that the imaginary part of the amplitude in \eqref{eq:dispB} cancels out from the lhs, leaving only its real part. Second, the definition of ${\cal B}(s,0^-)$ also includes an integral in the region $0\leq s \leq 4m^2$, which regulates IR divergences for massive external fields, ensuring finiteness of the physical result. Notice that when only massive modes are exchanged, our bounds reduce trivially to the standard bounds in the absence of massless poles.

From now on, all expressions can be equivalently used with either $\Sigma^{(j)}$ or $\hat \Sigma^{(j)}$, the only difference being the lower limit of the integral in the rhs of the dispersion relation. However, its explicit positivity does not change. Provided that the amplitude ${\cal A}(s,0^-)$ is finite, these bounds are applicable and can lead to interesting constraints on the structure of EFT Lagrangians through the presence of the Wilson coefficients in ${\rm Re}{\cal A}(s,0^-)$.

\section{Regularity in the forward limit}
Although the bounds \eqref{eq:positivity} and \eqref{eq:beyond} are completely general and valid in the case of massless particles in the spectrum of the theory, they are meaningless in the presence of divergences in the forward limit, as it is the case when gravitons are exchanged in the $t$-channel.

In that case, the lhs of the bound is dominated by the tree-level contribution to the exchange, which is of the form
\begin{align}
    {\cal A}(s,t)\propto -{\cal R}\times \frac{s^2}{t} ,
\end{align}
where ${\cal R}$ is the residue in the pole of the graviton propagator.

Since the limit $t\rightarrow 0^-$ is continuous --although divergent --, in principle we can always use \eqref{eq:positivity} to fix the sign of the divergence and conclude that
\begin{align}
    {\cal R}>0,
\end{align}
which tells us that in order to agree with unitarity requirements, the graviton must not be a ghost. Although it is interesting to see this trivial condition for unitarity arising in this way, the information that it provides is scarce. If we want to extract more information from the positivity bounds \eqref{eq:positivity} and \eqref{eq:beyond} in the presence of a graviton in the spectrum, then we need to find a way to regularize the forward limit divergences.

In \cite{Tokuda:2020mlf} it is shown\footnote{Note however that the bounds derived in \cite{Tokuda:2020mlf} do not take into account the presence of the branch cut at all.} that this is possible for the pole $t^{-1}$ if one takes a seemingly strong assumption about the scattering amplitude -- that it takes the Regge form \cite{Collins:1971ff} 
\begin{align}\label{eq:regge}
    {\rm Im}{\cal A}(s,t)=r(t)\left(\alpha's\right)^{2+l(t)}\left(1+\frac{\zeta}{\log(\alpha's)}+{\cal O}\left(\frac{1}{\alpha's}\right)\right),
\end{align}
which we extend here with a sub-leading correction, above a certain energy scale $E\sim M_*$. Here $r(t)$ encodes information about the polarization of external states, while $l(t)$ is constrained to be negative $l(t)<0$ and satisfies $l(0)=0$, in order for the amplitude to unitarize at high energies. The scale $\alpha'$ is controlled by the value of the Regge scale $M_*$ as $\alpha'\sim {\cal O}(1) M_*^{-2}$. Note that in \cite{Tokuda:2020mlf} the logarithmic correction that we include here is not considered. 

By assuming this behavior of the amplitude at large $s$, it can be easily shown that the rhs of \eqref{eq:sigmabar} will also present divergences when $t\rightarrow 0$, which are thus controlled only by the large $s$-limit of the integral. These can then be cancelled against those in the lhs, with the remaining finite piece satisfying its own version of positivity. Retaining only the leading term in the amplitude allows to cancel the pole $t^{-1}$ but leaves the logarithmic divergence untouched. As we will see in a moment, the sub-leading correction that we have included accounts for the latter. 

Of course, at this point one could question the validity of the high-energy behavior \eqref{eq:regge}. So far this is an assumption of our work, but one which is well-justified in the case of gravitons for two different reasons. First, let us give a heuristic argument. If we believe that String Theory provides a UV completion of gravitational interactions, then it can be shown that graviton mediated scattering amplitudes satisfy \eqref{eq:regge} at ${\cal O}(1)$, where $\alpha'$ is the string scale \cite{Camanho:2014apa,DAppollonio:2015fly,Hamada:2018dde}. This happens due to the contribution of the tower of massive modes in the spectrum that are excited above these energies. Logarithmic corrections of similar form to the ones in \eqref{eq:regge} can also be found in certain cases \cite{Camanho:2014apa} and we expect them to arise from string loops\footnote{Higher loop contributions like $\log(\log t)$ are expected beyond one-loop in the scattering amplitude. We expect them to cancel against higher loop corrections in the string theory.}. Second, if instead we assume an arbitrary sub-leading correction $g(s)$, it can be checked that the only choice that allows for cancelling the logarithmic divergence is precisely $g(s)=\zeta/\log(\alpha's)$, as it is shown in Appendix \ref{g_general}. Therefore, from now on we assume \eqref{eq:regge}. Note that by assuming \eqref{eq:regge}, the bound \eqref{eq:frm} is automatically satisfied.

We can then split the integral on the rhs of \eqref{eq:sigmabar} in two
\begin{align}
    \int_{\Lambda^2}^\infty dz\ F^{(j)}(z)=\int_{\Lambda^2}^{M_*^2} dz\ F^{(j)}(z)+\int_{M_*^2}^\infty dz\ F^{(j)}(z).
\end{align}

Calling $\Delta=\int_{M_*^2}^\infty dz\ F^{(j)}(z)$ and using the Regge form of the scattering amplitude \eqref{eq:regge}, we get
\begin{align}\label{eq:Delta}
    \Delta&=\frac{r(t)\alpha'^{2+l(t)}}{\pi}\int_{M_*^2}^\infty dz\ z^{3+l(t)-4j}\left(1+\frac{\zeta}{\log(\alpha'z)}\right),
\end{align}
which can be computed explicitly in terms of the analytic continuation of the Gamma function.

The forward limit can be taken in this expression after integration. Note that since $l(0)=0$, $t\rightarrow 0^-$, and $l(t)<0$, we have
\begin{align}\label{eq:l_expansion}
    l(t)=l'(0) t + \frac{l'' (0)}{2} t^2+{\cal O}(t^3),
\end{align}
with $l'(0)>0$. We thus get
\begin{widetext}
\begin{align}\label{eq:result_Delta}
\lim_{t\rightarrow 0^-}\Delta=\frac{r(0)\alpha'^2}{\pi}
\begin{cases}
\left(\frac{(M^2_*)^{4-4j}}{4j-4}+\zeta \alpha'^{4j-4}\Gamma\left[0,(4j-4)\log(M_*^2\alpha')\right]\right), & j>1\\
\left(\frac{1}{l'(0) t}-\frac{l''(0)}{2 l'(0)^2}+\log(M_*^2 \alpha')\right)-\zeta\left( \gamma+ \log(t)+\log\left[-l'(0)\log(M_*^2 \alpha')\right] \right), & j=1
\end{cases},
\end{align}
\end{widetext}
up to terms which vanish when $t=0$. Here $\gamma$ is the Euler-Mascheroni constant, $\Gamma(s,x)=\int_x^\infty dt\ t^{s-1} e^{-t}$ is the incomplete Gamma function, and we have taken $z\gg m^2,\delta$. We have also assumed that our external states satisfy ${\rm Im}{\cal A}^{\times}(s,t)={\rm Im}{\cal A}(s,t)$ from crossing symmetry, which limits the application of our result to bosonic states. Fermions will introduce extra signs from crossing symmetry.

We find that indeed the leading term in \eqref{eq:regge} produces a pole $t^{-1}$, while the sub-leading correction gives a $\log(t)$. However, note that they only exist when $j=1$, while for $j>1$ the result is completely regular. This isexactly the same kind of divergences that we find in the forward limit of $\Sigma^{(j)}$, only present for $j=1$ as well\footnote{The divergent part of the amplitude for graviton exchange is proportional to $s^2$. Thus, it vanishes from $\Sigma^{(j)}$ with $j>1$ after evaluation of the residues in the pole.}. Thus, we can expand both sides of expressions \eqref{eq:sigma} and \eqref{eq:sigmabar} in the limit $t\rightarrow 0^{-}$ and cancel divergences in the lhs against those in the rhs provided by $\Delta$, with the rest of the terms remaining finite. It is particularly interesting to note that assuming Regge behavior, which is expected to arise in gravity, precisely allows for cancellation of those divergences produced in graviton scattering. As discussed in Appendix \ref{g_general}, this seems to be a unique result.

Explicitly, using \eqref{eq:result_Delta} we can now rewrite \eqref{eq:sigmabar} for $j=1$ as
\begin{align}\label{eq:reg1}
    \nonumber \hat{\Sigma}_R^{(1)}&=\int_{\Lambda^2}^{M_*^2}dz\ F^{(1)}(z)+\frac{r(0)\alpha'^2 \log(M_*^2 \alpha')}{\pi}\\
    \nonumber &-\frac{r(0)\alpha'^2}{\pi}\frac{l''(0)}{2 l'(0)^2}-\frac{r(0)\alpha'^2 \zeta}{\pi}\log\left[-l'(0)\log(M_*^2 \alpha')\right]\\
    &-\frac{r(0)\alpha'^2 \zeta \gamma}{\pi} ,
\end{align}
where we have introduced the regularized version of $\hat{\Sigma}^{(1)}$ as
\begin{align}\label{eq:sigma_regu}
    \hat{\Sigma}_R^{(1)}=\hat{\Sigma}^{(1)}+\frac{r(0)\alpha'^2}{\pi}\left(\zeta \log(t)-\frac{1}{l'(0) t}\right),
\end{align}
by taking all divergent terms to the lhs.

By choosing the appropriate value of the combinations $r(0) \alpha'^2/l'(0)$ and $r(0)\alpha'^2 \zeta$, the forward limit divergences can be cancelled, so that \eqref{eq:reg1} remains regular. Note that, since $\alpha'^ 2>0$ and $l'(0)>0$ this also fixes the sign of $\zeta$ uniquely -- although in a case by case way.

Finally, we turn our attention to the explicit form of \eqref{eq:reg1}. Note that the integral along $\Lambda^2<s<M_*^2$ must remain positive by application of \eqref{eq:optical}. However, the rest of the terms do not have a definite sign. In particular, we cannot determine the overall sign of the rhs in \eqref{eq:reg1} without knowing the value of $l''(0)$, which we do not know. Nevertheless, all these terms come multiplied by the overall scale $r(0)\alpha'^2$. Thus, what we can do is to assess that the rhs is positive \emph{up to the order in which they become important}. Meaning
\begin{align}\label{eq:b1}
    \hat{\Sigma}_R^{(1)}>-{\cal O}(r(0)\alpha'^2),
\end{align}
so that a small amount of positivity violation is allowed and controlled by the dynamics of the UV degrees of freedom.

For $j>1$ things are simpler. Since \eqref{eq:result_Delta} is always convergent in this case, there is no need to explicit the divergence nor to split the range of integration in \eqref{eq:Delta}. Thus, we simply recover our result \eqref{eq:beyond}, which remains valid
\begin{align}\label{eq:b2}
    \hat{\Sigma}^{(j>1)}>0.
\end{align}

Expressions \eqref{eq:b1} and \eqref{eq:b2} are the final results of our work. They represent positivity bounds whose lhs can be computed in an EFT, as long as $\delta<\Lambda^2$, and whose value is constrained by features of the high-energy theory.

\section{Gravitating Scalar Field}
Now that we have derived useful positivity bounds in the presence of exchange of gravitons, let us test their validity with some well-known theories of scalar fields coupled to Einstein Gravity. The first case that we examine is a free gravitating scalar field, with action
\begin{align}
    S=\int d^4x\sqrt{|g|}\ \left(-\frac{R}{2\kappa^2}+\frac{1}{2}\partial_\mu \phi \partial^\m \phi\right),
\end{align}
where $\kappa^2=8\pi G=M_P^{-2}$.

In order to include the branch cut into the scattering amplitude $\phi\phi\rightarrow \phi\phi$ we must at least compute the first loop correction. Combining it with the tree level amplitude we get, in the forward limit and after renormalization\footnote{The coefficient in front of the logarithms is gauge dependent. However, its sign is universal for the family of $\beta$-gauges \cite{Barvinsky:1985an} explored here.}
\begin{align}
\nonumber    {\cal A}(s,0^-)&=-\frac{\kappa^2 s^2}{t}-\frac{33\kappa^4s^2}{24\pi^2} \left(\log(s)+\log(-s)\right)\\
    &-\frac{33\kappa^4s^2}{24\pi^2}\log(t).
\end{align}
Here we have used the de Donder gauge and set the renormalization scale $\mu_{\rm R}=1$ in the modified minimal subtraction scheme. This choice is harmless since its value always drops from the result.

The integral in \eqref{eq:defB} vanishes for massless external fields. Thus
\begin{align}
   \nonumber &{\cal B}(s,0^{-})={\rm Re}{\cal A}(s,0^-)=-\frac{\kappa^2 s^2}{t}-\frac{33\kappa^4s^2}{24\pi^2} \log(s^2)\\
    &-\frac{33\kappa^4s^2}{48\pi^2}\log(t^2),
\end{align}
and from this we can easily use \eqref{eq:sigma} to compute
\begin{align}
 &\Sigma^{(1)}=-\frac{\kappa^2}{t}-\frac{33 \kappa^4}{24\pi^2} \left(\frac{3}{2}+\log(t)+\log(\delta^2)\right),\\
&\Sigma^{(j>1)}=\frac{y(j) \kappa^2}{ \pi^2\delta^{4j-4}},
\end{align}
where $y(j)>0$ for all $j$. Here we have decided not to add the contribution from $\int_0^{\Lambda^2} dz F^{(j)}(z)$, thus working with $\Sigma^{(j)}$ instead of $\hat{\Sigma}^{(j)}$.

Cancelling the divergences using \eqref{eq:sigma_regu} determines  $r(0)\alpha'^2 \sim -l'(0)\kappa^2$ and $r(0)\alpha'^2 \zeta\sim  \kappa^4$. Thus the bounds read
\begin{align}
    &-\frac{33 \kappa^4}{24\pi^2} \left(\frac{3}{2}+\log(\delta^2)\right)> -{\cal O}(r(0)\alpha'^2),\\
    &\frac{y(j) \kappa^2}{ \pi^2\delta^{4j-4}}> 0.
\end{align}
The first bound is however meaningless since the lhs is already comparable to the sub-leading terms in the rhs. This forbids us to conclude anything from $\Sigma^{(1)}$. On the other hand, the second bound is automatically satisfied for all $\delta$, confirming a trivial statement, that a free gravitating scalar field is a \emph{bona fide} theory up to $M_P$.
\\
\section{Scalar QED}
Even more interesting is to explore the case of scalar QED with a photon $\phi$, an electron $\psi$ and an spectator field $\chi$, as suggested in \cite{Alberte:2020jsk}. The action is
\begin{align}\label{eq:UV}
\nonumber S&=\int d^4x\sqrt{|g|}\bigg[-\frac{R}{2\kappa^2}+\frac{1}{2}\partial_\m \phi \partial^\m \phi+\frac{1}{2}\partial_\m \chi \partial^\m \chi\\
&+\frac{1}{2}\partial_\m \psi \partial^\m \psi -\frac{1}{2}\Lambda^2\psi^2 -\lambda \Lambda \phi \psi^2\bigg].
\end{align}

At energies below the mass of the electron $\Lambda\ll M_P$, $\psi$ can be integrated out, leaving a generic EFT describing effective interactions between the rest of fields
\begin{align}\label{eq:EFT}
    \nonumber S&=\int d^4x\sqrt{|g|}\bigg[ -\frac{R}{2\kappa^2}+\frac{1}{2}\partial_\m \chi \partial^\m \chi + \frac{1}{2}\partial_\m \phi \partial^\m \phi \\
    \nonumber &-\frac{\lambda^3\Lambda}{(2\pi)^2}\frac{\phi^3}{3!}+\frac{\lambda^4}{2\pi^2}\frac{\phi^4}{4!}+\frac{D \lambda^2 \kappa^2}{\Lambda^2}(\partial\phi)^4\\
    &+\frac{C\lambda^2 \kappa^2}{\Lambda^2}(\partial_\m \phi \partial^\m \chi)^2+\dots\bigg],
\end{align}
where the dots indicate further $\Lambda$ or $\kappa^2$ suppressed terms. In matching both actions, the Wilson coefficients $D$ and $C$ must be determined by a direct comparison of a scattering amplitude. However here we are interested on exploring what positivity can say about them. Following \cite{Alberte:2020jsk} we focus on $\phi\chi\rightarrow \phi \chi$, whose one-loop amplitude gives
\begin{align}
    \nonumber {\cal A}(s,0^-)&=-\frac{\kappa^2 s^2}{t}+\frac{s^2 C\lambda^2 \kappa^2 }{\Lambda^2}- \frac{11 \kappa^4 s^2}{24\pi^2} \log(t)\\
    &- \frac{11 \kappa^4 s^2}{24\pi^2} \left(\log(s)+\log(-s)\right)+{\cal O}(t).
\end{align}

Again, we have added the one-loop correction, with $\mu_{\rm R}=1$, in order to make the branch cut explicit. From here we find
\begin{align}
   \nonumber  \Sigma^{(1)}&=\frac{\kappa^2}{48}\bigg(-\frac{48}{t}+\frac{48 C \lambda^2}{\Lambda^2}-\frac{33\kappa^2}{\pi^2}\\
   &-\frac{22\kappa^2}{\pi^2}\log(t)-\frac{22\kappa^2}{\pi^2}\log(\delta^2)\bigg),\\
    \Sigma^{(j>1)}&=\frac{y(j) \kappa^4}{\pi^2 \delta^{4j-4}}.
\end{align}

Cancelling the forward divergences we get again $r(0)\alpha'^2\sim -l'(0)\kappa^2$, $r(0)\alpha'^2 \zeta\sim  \kappa^4$. From this the bound $\Sigma^{(j>1)}>0$ is automatically satisfied. It also allows us to disregard the loop corrections in $\Sigma^{(1)}$, since they are sub-leading. We thus get
\begin{align}\label{eq:bound_qed}
  \frac{C \kappa^2 \lambda^2}{\Lambda^2}> -{\cal O}\left(r(0)\alpha'^2\right).
\end{align}

This result agrees with that of \cite{Bellazzini:2019xts,Alberte:2020jsk}, where it is proposed from different arguments. This also proves the conjecture in their conclusions of new physics required at a scale $(r(0)\alpha'^2)^{-1/4}<M_P$ in order to unitarize the theory. This can be seen from the fact that a direct matching between the EFT \eqref{eq:EFT} and its partial UV completion \eqref{eq:UV} demands $C<0$ with $C\sim {\cal O}(1)$, which violates our bound. Thus, \eqref{eq:UV} needs to be completed at intermediate energies.
\\
\section{Conclusions}
In this paper we have derived \emph{new} positivity bounds in the presence of exchange of massless particles between bosonic states. They generalize and \emph{formalize} previous results in the literature. Provided that divergences in the forward limit can be ignored, our bounds can constrain the value of Wilson coefficients and other couplings in EFTs for which the existence of a plausible unitary, Lorentz invariant, and local UV completion is demanded.

We have gone further and shown that in the case of exchange of gravitons, forward divergences can be cancelled by assuming a Regge behavior of the scattering amplitude, which is unique if one assumes analyticity of the function $l(t)$. Although peculiar, this form of the ampplitude has been previously found in the literature in String Theory. This leads to well-defined bounds which can now be used in the presence of gravity. 

We have shown how our bounds work in two simple examples. A free gravitating scalar field, where they are automatically satisfied; and scalar QED with a spectator field, for which they demand new physics below the Planck scale to unitarize the theory, as previously suggested by \cite{Bellazzini:2019xts,Alberte:2020jsk}.

These new bounds open up a window to explore the theory space of phenomenological viable theories of (matter and) gravity. We believe that our results here have the potential to highly constrain different popular models currently used to investigate properties of black hole physics and Cosmology. It would also be interesting to apply them to the exploration of unitarization mechanisms for graviton scattering \cite{Blas:2020dyg,Blas:2020och,Aydemir:2012nz}.

\section*{Acknowledgements}
We are grateful to Brando Bellazzini, Javi Serra and Inar Timiryasov for discussions and comments. Our work has been supported by the European Union's H2020 ERC Consolidator Grant “GRavity from Astrophysical to Microscopic Scales” grant agreement no. GRAMS-815673 (M. H-V.), by the Spanish FPU Grant No FPU16/01595 (R. S-G.) and by the Academy of Finland grant 318319 (A. T.). The part of work of A. T. related to obtaining the bounds from imaginary poles was supported by the Russian Science Foundation grant 19-12-00393. We also wish to acknowledge networking support from COST action CA16104 ``GWverse".

\appendix
\section{Universality of the sub-leading correction}\label{g_general}
Let us address here the question on the uniqueness of the sub-leading correction to the amplitude in the Regge limit \eqref{eq:regge} required to cancel divergences in the forward limit.

Let us start by noticing again that the leading term, which cancels the $t^{-1}$ contribution of the amplitude in the IR, was already proposed in earlier works \cite{Tokuda:2020mlf} and can be obtained from a closed string amplitude after a careful manipulation. In particular, the imaginary part of the string amplitude is not a regular function of $s$ and $t$, it has instead an infinite set of Regge poles that require regularization. Hereinafter we will assume instead that the imaginary part of the Regge amplitude that we consider is regular in both arguments when $s\rightarrow\infty$, $t\rightarrow 0$. 

Going back to $F^{(j)}(z,t)$, defined in \eqref{rhs}, let us examine the integral in \eqref{eq:Delta}
\begin{align}
    \Delta_j=\int_{M_*^2}^{\infty} dz\ F^{(j)}(z,t).
\end{align}
Note that this integral can give a singularity at $t\rightarrow 0$ only if it is divergent when $t=0$ but finite for some small finite $t$. In particular, for $j=1$ we obtain
\begin{align}
    \Delta_1 = \int_{M_*^2}^{\infty} dz\ \frac{{\rm Im} {\cal A}(z,t)}{z^3}
\end{align}

Since ${\rm Im} {\cal A}(s,t)$ is regular at $t=0$ from our assumption, we can Taylor expand it around this point
\begin{align}
    {\rm Im} {\cal A}(s,t)={\rm Im}{\cal A}(s,0)+\left. \partial_t{\rm Im}{\cal A}(s,t)\right|_{t=0} t +{\cal O}(t^2),
\end{align}
in one to one correspondence to the series expansion of the Regge form \eqref{eq:regge},
\begin{align}
    \nonumber &{\rm Im} {\cal A}(s,t)=r(t)(\alpha' s)^{2+l(t)}\left(1+g(s)\right)\sim 
    \\
    &=(\alpha' s)^{2}\left(1+g(s)\right)\left[r(0)+t(r'(0)-l'(0)\log(s)\right],
\end{align}
where we have assumed the expansion \eqref{eq:l_expansion}.

At this point we leave the form of the sub-leading correction $g(s)$ completely arbitrary. If we demand that the result of $\Delta_1$ has the correct divergent structure we have
\begin{align}
    \Delta_1 = \int_{M_*^2}^{\infty} \frac{dz}{z}r(t)z^{-l(t)}(1+g(z))=\frac{a}{t}+b(t)+{\cal O}(1).
\end{align}
Here $b(t)$ stands for the remaining divergent terms at $t\rightarrow 0$, which include the one-loop $\log t$ term among others. Changing the integration variable to $\log z=\sigma$ and plugging the small $t$ expansion on the integrand, brings us to the condition
\begin{align}
    \nonumber &r(t)\int_{\log{M_*^2}}^{\infty}d\sigma e^{-(l'(0)t+{\cal O}(t^2))\sigma}(1+g(\sigma))\\
    &=\frac{a}{t}+b(t)+{\cal O}(1).
\end{align}

The leading term in the left hand side can be computed explicitly and shown to cancel the $a t^{-1}$ term, while for the rest we have
\begin{align}
    r(0)\int_{\log{M_*^2}}^{\infty}d\sigma e^{-(l'(0)t)\sigma}g(\sigma) =b(t)+{\cal O}(1).
\end{align}

After multiplying by a step function under the integral sign, this becomes a Laplace transform. Although it requires regularization, its result is unique and therefore there exists a single function $g(s)$ which satisfies this identity. Since $g(s)=\zeta/\log (\alpha's)$ does the work, we conclude that it is the only option.

\bibliography{positivity}{}

\end{document}